\documentclass[11pt]{article}

\usepackage{arxiv}
\usepackage{amssymb,amsmath,amsthm}

\usepackage{bm}
\usepackage[colorlinks,citecolor=blue]{hyperref}
\usepackage{subcaption}
\usepackage{multirow}
\usepackage{mathtools}
\usepackage{threeparttable}
\usepackage{booktabs}
\usepackage{blkarray, bigstrut} %
\usepackage{arydshln}
\usepackage{rotating}
\usepackage[numbers]{natbib}
\newcommand{\indep}{\perp \!\!\! \perp}

\begin{document}
\title{Marginal Structural Models with Latent Class Growth Modeling of Treatment Trajectories}
\author{Awa Diop \\
     awa.diop.2@ulaval.ca \\
     D\'epartement de m\'edecine sociale et pr\'eventive \\	Universit\'e Laval 
   \And
     Caroline Sirois \\
  Caroline.Sirois@pha.ulaval.ca \\
  Facult\'e de pharmacie\\	Universit\'e Laval 
	\And
	  Jason Robert Guertin \\
    jason.guertin@fmed.ulaval.ca\\
    D\'epartement de m\'edecine sociale et pr\'eventive \\	Universit\'e Laval 
			\And
    Denis Talbot \\
    denis.talbot@fmed.ulaval.ca \\
    D\'epartement de m\'edecine sociale et pr\'eventive \\	Universit\'e Laval 
}

\maketitle

\begin{center}
``The final, definitive version of this paper has been published in the Journal Statistical Methods in Medical Research. DOI: https://doi.org/10.1177/09622802231202384. \textcopyright[Diop, A., Sirois, C., Guertin, J.R., Schnitzer, M.E., Candas, B., Cossette, B., Poirier, P., Brophy, J., Mésidor, M., Blais, C. and Hamel, D., 2023]."
\end{center}

\begin{abstract}

{In a real-life setting, little is known regarding the effectiveness of statins for primary prevention among older adults, and analysis of observational data can add crucial information on the benefits of actual patterns of use. Latent class growth models (LCGM) are increasingly proposed as a solution to summarize the observed longitudinal treatment in a few distinct groups. When combined with standard approaches like Cox proportional hazards models, LCGM can fail to control time-dependent confounding bias because of time-varying covariates that have a double role of confounders and mediators. We propose to use LCGM to classify individuals into a few latent classes based on their medication adherence pattern, then choose a working marginal structural model (MSM) that relates the outcome to these groups. The parameter of interest is nonparametrically defined as the projection of the true MSM onto the chosen working model. The combination of LCGM with MSM is a convenient way to describe treatment adherence and can effectively control time-dependent confounding. Simulation studies were used to illustrate our approach and compare it with unadjusted, baseline covariates-adjusted, time-varying covariates adjusted and inverse probability of trajectory groups weighting adjusted models. We found that our proposed approach yielded estimators with little or no bias. } 

\end{abstract}

\quad

\begin{center}
\end{center}

\clearpage
\section{Introduction}

Cardiovascular disease (CVD) is the leading cause of death worldwide  (\citep[]{WHO1}).  A well known treatment to prevent  CVDs is HMG-CoA reductase inhibitors (statins). Several randomized controlled trials (RCTs) have shown the efficacy of statins to prevent a first event of CVD, that is, for primary prevention (\citep[]{Taylor11}). Despite current evidence, it is still not clear if these conclusions can be applied to older adults as they are often excluded from RCTs (\citep[]{singh2018statins,armitage2019efficacy}). Also, treatment adherence is known to be lower in real-world situations  than in RCTs (\citep{chowdhury2013adherence}). Hence, there is little evidence concerning the effectiveness of statins for primary prevention among older adults in a real-life setting. Analysis of observational data could add crucial information on the benefits of actual statin's patterns of use among older adults.  However, the number of unique treatment trajectories increases exponentially with the length of follow-up in longitudinal studies.  

Latent class growth models (LCGM), or group-based trajectory models, are increasingly being proposed as a solution to summarize the observed trajectories in a few distinct groups (\citep[]{nagin2005group,franklin2013group}). LCGM defines homogeneous subgroups of individuals with respect to their patterns of change over time (\citep[]{nagin2005group,muthen2001latent}). Recently \citep{franklin2013group} applied LCGM to classify patients who initiated a statin according to their long-term adherence. The authors compared LCGM with traditional approaches to measure medication adherence such as the proportion of days covered or medication possession ratio. They concluded that LCGM is a better method to measure adherence and may facilitate targeting of interventions as it made describing adherence behaviors easier.  \citep{franklin2015association} have combined LCGM and Cox models to describe the association between statin adherence and cardiovascular events. 
 
 While LCGM is a promising approach for estimating the effect of treatment trajectories in real-life settings, adjustment for potential confounders is essential. Indeed, time-dependent confounders are susceptible to be affected by previous treatments (\citep{3hernan2000marginal}). As a consequence, time-varying covariates can have a double-role of confounders and mediators. In such situations, standard approaches, such as covariate adjustment in an outcome regression model, fail to control confounding  (\citep{3hernan2000marginal}). Marginal structural models (MSMs) are well known in the causal inference literature for their ability to deal correctly with time-varying covariates and time-varying exposure. 
 
Estimating the parameters of a MSM that links the outcome to all possible treatment trajectories over long follow-up periods is unlikely to be realistic in most applications. A common alternative is to model the outcome as a function of the cumulative number of periods of treatment adherence. However this approach lacks flexibility since it assumes a linear relationship between cumulative treatment and outcome. A flexible Cox MSM with such weighted cumulative exposure was proposed by \citep{xiao2014flexible}. The weighted cumulative exposure is estimated with splines allowing a data-driven specification of the relation between the outcome and the treatment trajectory. While this alternative is flexible, it is not as easy to describe treatment adherence with it. An attempt to combine LCGM with MSM was realized by \citep{lim2015all}  but the authors did not use an appropriate estimation method to control for time-varying confounders. Their method is similar to combining LCGM with standard modeling approachs.

In this paper, we propose a suitable theoretical framework to combine LCGM and MSMs. We first introduce a finite mixture modeling approach (LCGM) to classify individuals into a few latent classes. Our approach consists in choosing a working MSM that relates the outcome to these latent classes.  The parameter of interest is nonparametrically defined as the projection of the true MSM onto the chosen working causal model (\citep[]{neugebauer2003locally}). Hence, we alleviate the burden of assuming a correctly specified model. As far as we know, such a combination of MSMs with LCGM that allows to control for time-varying confounders, has never been investigated. In the remainder, we present the modeling framework (notation, causal model, mixture model, estimation approach) followed by a simulation study to illustrate and investigate the finite-sample properties of our proposal. We end this paper with some recommendations for the practical application of our proposed approach.
\section{Notation and nonparametric marginal structural model definition}
\label{sec2}
In the following, we use capital letters to represent random variables and corresponding lower case letters to represent observed or fixed values these variables take. Consider a longitudinal study where $t=1, \ldots , K$ is the follow-up time. Let $\bar{A}_{t}=(A_{1} , A_{2},\ldots,A_{t})$ be the treatment trajectory up to time $t$. Similarly, $\bar{L}_t = (L_1, L_2, ... L_t)$ is the covariates' history up to time $t$.  These covariates are assumed to include all baseline and time-varying confounders. As a notational shortcut, we define $\bar{A} \equiv \bar{A}_K$ and $\bar{L} \equiv \bar{L}_K$.  Let $Y$ be the observed outcome at the end of follow-up. We denote by $Y^{\bar{a}}$ the counterfactual outcome under a specific treatment trajectory, that is, the outcome that would have been observed had the treatment trajectory been $\bar{a}$, possibly contrary to fact. Similarly, $\bar{L}^{\bar{a}}$ denotes a counterfactual covariate process. We denote by $F_{X}$ the unknown distribution of all possible counterfactual variables $X=(\bar{L}^{\bar{a}},Y^{\bar{a}}: {\bar{a} \in \mathcal{A}})$  where $\mathcal{A}$ is the set of all possible values of $\bar{a}$. We denote by $P_{F_X}$ the distribution of the observed data $\mathcal{O}=\{Y, \bar{A}_{K},\bar{L}_{K}\}$, from which $i = 1,...,n$ independent and identically distributed observations are randomly drawn. We consider the following nonparametric form of a MSM where we define $V$ as a (possibly empty) subset of the baseline covariates, $V \subseteq L_1$: 
\begin{equation}
E_{F_X}(Y^{\bar{a}}|V) = m^{*}(\bar{a}, V).
\label{msmgen}
\end{equation}
Equation (\ref{msmgen}) represents a model where the distribution of all possible counterfactual variables, $F_X$, is left unknown. Thus, we do not make the assumption of a correctly specified model as would be the case when considering a parametric MSM. The nonparametric identification of model (\ref{msmgen}) from the observed data can be achieved under the following causal assumptions: 1) sequential conditional exchangeability,  2) no interaction between subjects, 3) consistency, and 4) positivity. Assumption 1) indicates that conditional on the history of treatment up to time $t-1$ and the history of the covariates up to time $t$ there are no unmeasured confounders (\citep[]{robins2000marginal}), that is $ Y^{\bar{a}} \bm{\indep} A_t | \bar{A}_{t-1},\bar{L}_{t}$.  Assumption 2) means that the potential outcome of a given individual is not affected by others' exposure. Assumption 3) means that, given the observed treatment trajectory, the observed outcome and the potential outcome under that given trajectory are the same, formally if $\bar{A} = \bar{a}$ then $Y^{\bar{a}}= Y$. Assumption 4) requires that for each level $(\bar{a}_t, \bar{l}_{t})$, $P(A_t = a_t |\bar{A}_{t-1} = \bar{a}_{t-1}, \bar{L}_{t} = \bar{l}_{t}) > 0$. In other words, in each stratum defined by previous treatment and covariates, we find both exposed and unexposed individuals at time $t$.
Under these assumptions, we have:
\begin{align}
\mathbb{E}_{F_X}\left(Y^{\bar{a}}\right) = \int_{\bm{l_1}} \ldots   \int_{\bm{l_{K}}}\mathbb{E}(Y|\bar{a}_{K}, \bar{\bm{l}}_{K}) f(\bm{l_{K}|\bm{l}_{K-1}},\bar{a}_{K-1} )\mu(\bm{L_{K}}), \ldots f(\bm{l_1})\mu(\bm{L_1}),
\label{eq.np}
\end{align} 

\noindent where $f(\bm{l_{K}|\bm{l}_{K-1}},\bar{a}_{K-1} ), \ldots f(\bm{l_1})$ are the densities of $\bm{\bar{L}_t}$ relative to appropriate measures $\mu(\bm{L_t}),t=1,\ldots,K$, respectively.
Due to the curse of dimensionality, Equation (\ref{eq.np}) is rarely estimable with finite sample data. As mentioned in the introduction, one challenge is related to the number of treatment trajectories. In the next two sections, we introduce the LCGM and our LCGM-MSM as a solution to this specific challenge. 

\section{Summarizing treatment trajectories with latent class growth models}\label{sec3}
We briefly review LCGM below, more details can be found elsewhere (for example, see \citep{nagin2005group}). The first step of our LCGM-MSM proposal consists in summarizing treatment trajectories with a LCGM. Let $C_i$ denote the latent group-membership or trajectory group for the $i^{th}$ individual estimated from the individual's observed treatment $\bar{A}_{i}, i=1,\ldots, n$. We denote by $\pi_j$ the probability that an individual belongs to the $j^{th}$ trajectory group in the population, $j=1,\ldots, J$ where $J$ is the total number of  groups with the constraints $\pi_j\geq 0$ and $\sum_{j=1}^J\pi_j=1$. In LCGM, a trajectory is often described by a polynomial function of time. For example, if a quadratic relation was assumed for the $j^{th}$ group, we could use the regression model $logit(P(\bar{A}_{it}|C_i = j)) = \theta_0^j + \theta_1^j t + \theta_2^j  t^2$ where the parameters $\theta^{j}, j=1,\ldots, J$ describe the treatment trajectory over time for the $j^{th}$ group. This specification implies that the following mixture model is the marginal probability of the treatment trajectories $P(\bar{A}_i)$:
\begin{equation}
P(\bar{A}_{i})=\sum_{j=1}^J\pi_j \prod_{t=1}^K P(A_{it}|C_i=j).
\label{mixt}
\end{equation}

\begin{figure}[h!]
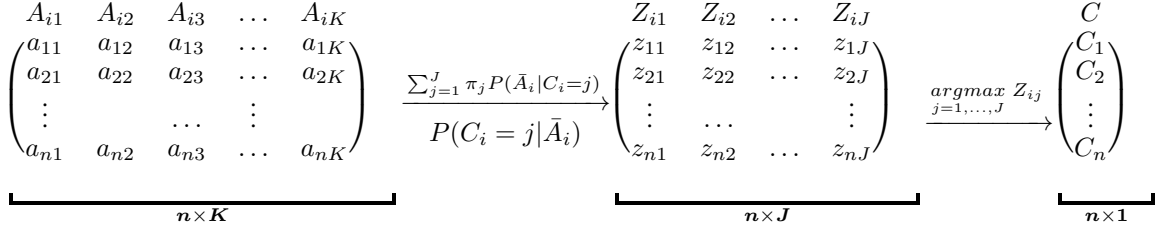

\centering
 \[\
 \underbracket{\begin{blockarray}{cccccc}
A_{i1} &  A_{i2}&  A_{i3} & \ldots &  A_{iK}\\
\begin{block}{(ccccc)c}
a_{11} & a_{12}& a_{13} &\ldots & a_{1K} & \\
 a_{21} & a_{22} & a_{23} &\ldots & a_{2K} &  \\
 \vdots& &\ldots &\vdots && \\
a_{n1}& a_{n2}& a_{n3}&\ldots & a_{nK} &  \\
\end{block}
\end{blockarray}}_{\bm{n}\times \bm{K}}
  \begin{matrix} \\ \xrightarrow{\sum_{j=1}^J\pi_jP(\bar{A}_{i}|C_i=j)} \\ P(C_{i}=j|\bar{A}_{i})\end{matrix}
 \underbracket{\begin{blockarray}{ccccc}
Z_{i1}& Z_{i2}&  \ldots &  Z_{iJ} \\
\begin{block}{(cccc)c}
z_{11} & z_{12}& \ldots & z_{1J}&\\ 
z_{21} & z_{22}& \ldots & z_{2J}&\\ 
\vdots&\ldots& &\vdots&\\ 
z_{n1} & z_{n2}& \ldots & z_{nJ}&\\ 
\end{block}
\end{blockarray}}_{\bm{n}\times\bm{J}}
  \begin{matrix} \\ \xrightarrow{\underset{j=1,\ldots, J} {argmax } \ Z_{ij}} \end{matrix}
   \underbracket{\begin{blockarray}{cc}
C & \\
\begin{block}{(c)c}
C_{1} & \\ 
C_{2} & \\ 
\vdots &\\ 
C_{n} \\ 
\end{block}
\end{blockarray}}_{\bm{n}\times\bm{1}}
\]

\caption{Summarizing treatment trajectories with LCGM.The matrix $n\times K$  represents treatment values for all the $n$ individuals and $K$ follow-up time, the matrix $n\times J$ represents matrix of $n$ individuals and their conditional probabilities, the vector $n\times 1$ represents the trajectory groups. The formulas on the arrows indicate how to obtain these matrices.}
\label{figLCGM}

\end{figure}

Model (\ref{mixt}) relies on the assumption of local independence, that is treatment uses at different time-points $A_{i1},\ldots, A_{iK}$ are assumed to be mutually independent conditional on group-membership (\citep[]{vermunt2004local}): $A_{it}\indep A_{it^{'}}|C_i=j, \ \text{for all} \  t\neq t^{'}, j=1,\ldots,J$. The log-likelihood of the $n$ observations is given by:
\begin{align*}
\mathcal{L}(\pi,\theta) &=\sum_{i=1}^nlog\left(\sum_{j=1}^J\pi_j \prod_{t=1}^K P(A_{it}|C_i=j,\theta^{j})\right).
\end{align*}
A general framework for fitting LCGM is implemented in the R package \textit{FlexMix} which is based on an Expectation-Maximization algorithm (\citep[]{leisch2004flexmix}). We denote by $Z_{ij}=P(C_{i}=j|\bar{A}_{i})$, the $i^{th}$ individual's conditional (``posterior") probability of membership in group $j$. These probabilities are computed postmodel estimation and are used to assign subjects into a trajectory group, usually based on their largest value (\citep[]{nagin2005group}). Therefore, we can formally define the trajectory groups as: $C_i  =\underset{j=1,\ldots ,J} {argmax } \ Z_{ij}$.  The conditional probability is given by:
\begin{equation}
P(C_i=j|\bar{A}_{i})=\dfrac{\pi_{j^{'}}P(\bar{A}_{i}|C_i=j^{'})}{\sum_{j=1}^J\pi_j P(\bar{A}_{i}|C_i=j)}.
\end{equation}
 In Figure (\ref{figLCGM}), we present a global overview of the LCGM approach. From the matrix $n\times K$ of treatment trajectories, we determine the $n\times J$ matrix of individual conditional probabilities which in turn yields the vector $n\times 1$ of trajectory groups. The specification of the order of the polynomial equation and the number of trajectory groups to include in the model are discussed in \citep{nagin2005group}. Note that the local independence assumption of the LCGM is unlikely to hold in presence of time-dependent confounding. Indeed, time-dependent confounding is susceptible to create a complex dependence structure between treatment uses at different time-points. If this assumption does not hold, LCGM can still be regarded as a convenient statistical technique to summarize treatment trajectories.

\section{Defining the causal parameter of interest}\label{sec4}
Once the observed treatment trajectories $\bar{A}$ are summarized using the LCGM, the next step of our LCGM-MSM approach is to formally define the causal effect of interest. We propose to approximate the nonparametric MSM $m^{*}(\bar{a}, V)$ for all $\bar{a} \in \mathcal{A}$ with a working MSM that is a function of the trajectory groups. Following a similar notation to \citep{neugebauer2007nonparametric},  we denote by $\mathcal{M}^*$ the infinite-dimensional space of all possible functions $m^*$ relating the counterfactual expectation with the treatment trajectories and baseline covariates $V$ as in Equation (\ref{msmgen}). We define $\mathcal{M}$ a subset of $\mathcal{M^*}$, consisting of all the the family of functions $m$ that the user is interested in using to characterize how the counterfactual expectations are related to the treatment trajectories and baseline covariates $V$ (i.e., the working model).   Our MSM-LCGM approach entails specifying $\mathcal{M} = \{m(C,V|\beta): \beta \in \mathcal{R}^p \}$, where $p$ is the number of parameters in $m(C,V|\beta)$. In other words, we are interested in characterizing the counterfactual outcomes as a function of the LCGM trajectory groups, for example $\mathbb{E}[Y|C,V] = \beta_0 + \beta_1 C_1 + ... + \beta_{J-1} C_{J-1}$. The model could feature additional parameters for covariates $V$ and for interactions terms between $C$ and $V$ if effect modification is of interest.  Remark that if $V = \emptyset$ and $J = 2^K$, then $\mathcal{M} = \mathcal{M}^*$. Indeed, the number of LCGM trajectory groups would then correspond with the total number of possible treatment trajectories, and the working model would be saturated.

Our causal parameter of interest is defined as a projection of the true nonparametric MSM $m^*$ on the working model $m$. This projection is notably characterized by a weight function $\lambda(\bar{A}, V)$ that determines, for all $(\bar{a}, v) \in \mathcal{A} \times \mathcal{S}_V$ where $ \mathcal{S}_V$ is the set of all possible values of $V$ (its support), how well the working MSM $m$ should approximate the true MSM $m^*$ relative to other values of $(\bar{a}, v)$. More formally:
\begin{align}
&\beta(\cdot|m,\lambda) : \mathcal{M^{*}}\rightarrow\mathcal{R}^{p}\\ \nonumber
&\beta_{m,\lambda}=\underset{\beta \in \mathbf{R}^p} {argmin} \lVert m^{*}-m(C,V|\beta)\rVert_{\lambda}.
\end{align}

 In other words, $\beta \equiv \beta_{m,\lambda} = \beta(F_X|m, \lambda)$ is chosen according to $\lambda$ to minimize the ``distance'' between the true model $m^*$ and the working model $m$ according to a given loss-function  $|| \cdot ||_{\lambda}$. When $Y$ is part of an exponential family, a natural choice for the loss function is derived from its log-likelihood (for more details see appendix).  Note that the trajectory groups $C$ are only a function of the time-varying treatment ($\bar{A}$) in the LCGM approach. Therefore, we can write $m(C,V|\beta) \equiv m(\bar{a},V|\beta)$. For example, when the outcome $Y$ is continuous, a weighted $L_2$ loss function is commonly used:
\begin{align*}
&\beta(\cdot|m,\lambda) : \mathcal{M}\rightarrow\mathcal{R}^{p}\\
&\beta(F_X|m,\lambda)=\underset{\beta \in \mathbf{R}^k} {argmin} \text{ } E_{F_X} \left[\underset{\bar{a} \in \mathcal{A}} \sum\left( Y^{\bar{a}}-m(\bar{a},V|\beta)\right)^2\lambda(\bar{a},V)\right].
\label{Ycontinu}
\end{align*}

The choice of $\lambda(\bar{a}, V)$ depends on the modeling goals. Common choices notably include $\lambda(\bar{a}, V) = 1$, which would be suitable if it is desired to get a global summary of the nonparametric causal model, and $\lambda(\bar{a}, V) = \prod_{t=1}^KP(A_t|\bar{A}_{t-1},V)$ which gives more weights to trajectories that are frequently observed. 

\textbf{Case of a survival outcome}\\
When $Y$ is a time-to-event outcome, we define our working causal model as a Cox proportional hazard MSM where $h_0(t)$ is the baseline hazard function:
\begin{equation}
h(t|\bar{a},V,\beta)=h_0(t)exp(\beta_1C_1+\ldots+\beta_{J-1}C_{J-1}).
\end{equation}
Indeed, $E(T^{\bar{a}}) = \int_t S(t) = \int_t exp(-H(t)) = \int_t exp(- \int_k^t h(t|\bar{a},V,\beta)) \equiv m(\bar{a},V|\beta)$.
 As in our illustration, the loss function can be defined from the negative partial log-likelihood. Let $Y^{\bar{a}}$ denote the minimum between the counterfactual survival time  $T^{\bar{a}}$ and the censoring time denoted by $U^{\bar{a}}$. The indicator for the event is designated by $\delta$ where $\delta=1$ if $T^{\bar{a}} < U^{\bar{a}}$ and 0 otherwise and $\mathcal{RS}(l)=\{i=1,\ldots n : T^{\bar{a}}>l\}$ designate the set of individuals at risk at time $l$. We defined the partial likelihood as a function of the groups as follows:
\begin{align*}
L(\beta)&=\prod_{i=1}^n\left\{\dfrac{h(t_i|\bar{a}_i, V_i,\beta) }{\sum_{l \in \mathcal{RS}(l)}h(t_l|\bar{a}_l, V_l,\beta) }\right\}^{\delta_i} \\
&= \prod_{i=1}^n\left\{\dfrac{exp(C_i\beta)}{\sum_{l \in \mathcal{RS}(l)}exp(C_l\beta)}\right\}^{\delta_i}
\end{align*}

Thus, the partial log-likelihood is given by:
\begin{align*}
log(L(\beta))&=\sum_{i=1}^n\delta_i\left\{log(h(t_i|\bar{a}_i, V_i,\beta))-log(\sum_{l \in \mathcal{RS}(l)}h(t_l|\bar{a}_l, V_l,\beta))\right\}\\
&=\sum_{i=1}^n I(T_i^{\bar{a}} < U_i^{\bar{a}})log\left\{\sum_{l \in \mathcal{RS}(l)}exp(C_i\beta-C_l\beta)\right\}
\end{align*}

 We define our parameter of interest $\beta$ as follows: 
\begin{align*}
&\beta(\cdot|m,\lambda) : \mathcal{M}\rightarrow\mathcal{R}^{p}\\
&\beta(F_X|m,\lambda)=\underset{\beta \in \mathbf{R}^k} {argmin} \text{ } E_{F_X} \left\{\bm{-}\sum_{\bar{a} \in \mathcal{\bar{A}}} \left[ log(L(\beta))\right]\lambda(\bar{a},V)\right\}.
\end{align*}

\section{Estimation of the parameter of the LCGM-MSM}\label{sec5}
 
We now describe how the parameter $\beta$ of our MSM-LCGM can be estimated from the observed data.  Because, the distribution of the trajectory groups $C$ does not depend on the parameter of interest $\beta$, $C$ is an ancillary statistic (\citep[]{basu1959family, lehmann1992ancillarity}).  Therefore, in the second-step of our approach, the trajectory groups are not seen as random but rather as a fixed regressor (\citep[]{lehmann2006testing}). In other words, the fact that the trajectories are estimated from the data can be ignored when estimating the parameter of our LCGM-MSM and when estimating the standard error of the estimator and confidence intervals. However, it should be noted that inferences are conditional on the selected LCGM.

\textbf{Inverse probability of treatment weighting}

Several approaches exist to estimate the parameters of a MSM and one of the most popular is the inverse probability of treatment weighting (IPTW). The IPTW mimics a random assignment of the treatment by creating a pseudo-population where the treated and the untreated groups are comparable. In longitudinal settings, IPTW can appropriately adjust for time-varying covariates affected by prior exposure and selection bias. Following \citep{neugebauer2003locally}, the IPTW is given by:
\begin{equation}
W(\bar{A},V)  =\dfrac{\lambda(\bar{A},V)}{g(\bar{A}| X)}
\end{equation}
\noindent where $g(\bar{A}|X)$ is the treatment mechanism which may depend on the counterfactual data $X$. However, under the sequential conditional exchangeability assumption, $g(\bar{A}|X)=\prod_{t=1}^K P(A_{t}=1|\bar{A}_{t-1},\bar{L}_{t})$. In the literature, stabilized weights (i.e when we set $\lambda(\bar{A},V)=\prod_{t=1}^KP(\bar{A}_t|\bar{A}_{t-1},V))$ are largely recommended since they tend to yield less extreme values (\citep[]{talbot2015cautionary}). In other words, each subject receives a weight that is proportional to the product of the inverse of the probability of their observed treatment at each time-point conditional on previous confounders and treatments. Note that the weights are thus computed based on the observed treatment trajectories ($\bar{A}$) and not on the trajectory groups $C$.

\textbf{IPTW estimating function}\\

\begin{figure}[h!]
\centering
\includegraphics[scale=0.75]{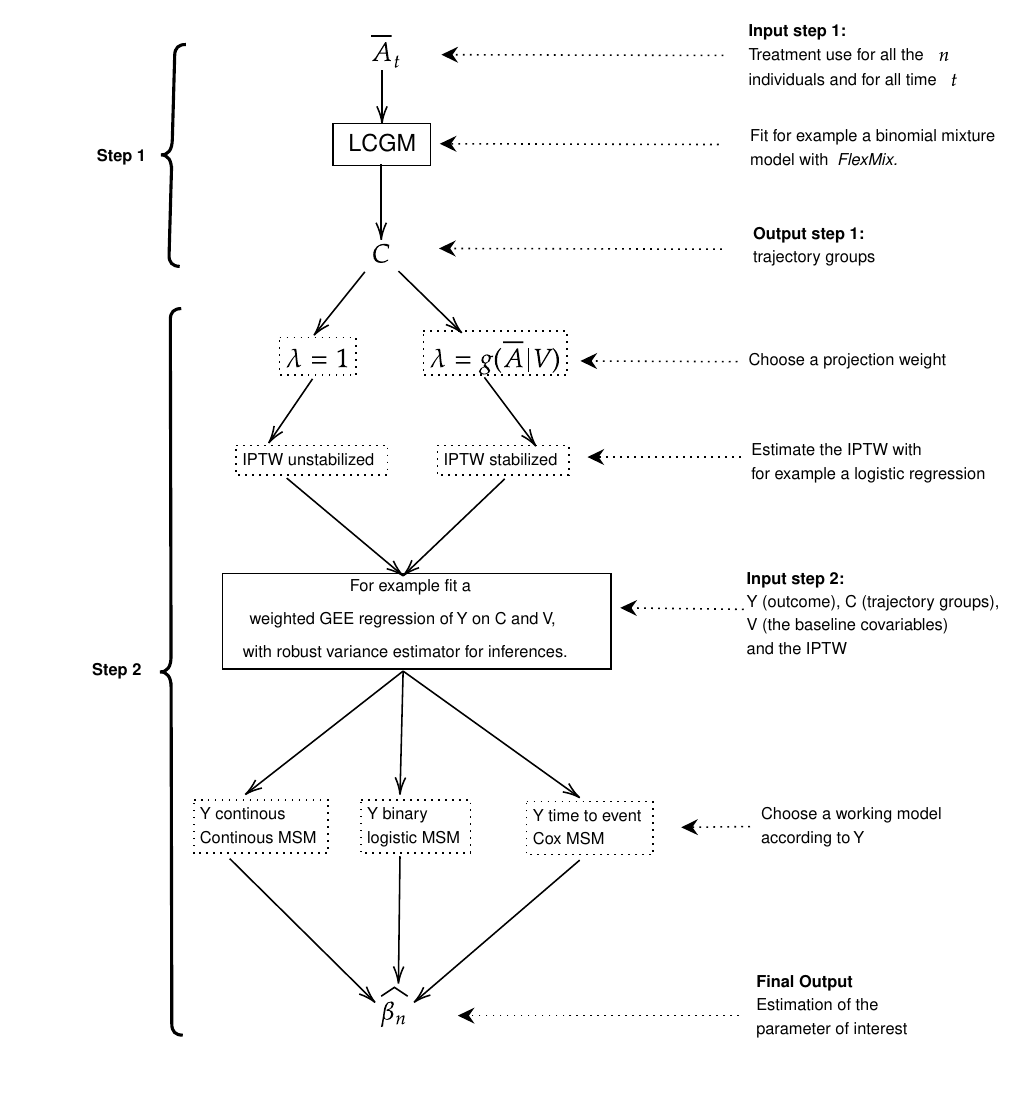}
\caption{Practical template for implementation of the LCGM-MSM approach}
\label{diagLCGMSM2}
\end{figure}

To estimate the parameter of interest $\beta_{m,\lambda}$, we have to define the estimating function based on the IPTW. We first differentiate the loss function and then we derive from its expression the IPTW estimating function. We denote by $D(X|\beta)$ its first order derivative. The IPTW estimating function  when $Y$ is continuous is given by:
 \begin{equation}
D(\mathcal{O}|\beta,g)= \dfrac{D(X|\beta)}{g(\bar{A}|X)}
\label{eq.est}
\end{equation}

\noindent where the first order derivative of the $L_2$ loss-function is given by: 
\begin{align*}
D(X|\beta)=\underset{\bar{a} \in \mathcal{A}} \sum \lambda(\bar{a},V)\dfrac{\partial}{\partial \beta}m(\bar{a},V|\beta)\epsilon_{\bar{a}}(\beta)
\end{align*}
  \noindent with $\epsilon_{\bar{a}}(\beta)=Y^{\bar{a}}-m(\bar{a},V|\beta)$ is the residual after projecting the nonparametric MSM onto the chosen working model. A well known property of the IPTW estimating function is its unbiasness at its true parameter which satisfy $E(D(\mathcal{O}|g,\beta))=0$ (see for example \citep[]{neugebauer2003locally}). Thus, it can be shown under the positivity assumption, that the IPTW estimator $\hat{\beta}_{m_{IPTW},\lambda}$ of $\beta_{m,\lambda}$ is the solution of the estimating equation $\sum_{i=1}^n D(\mathcal{O}_i|g_n,\beta)=0$ where $\lambda_n$ and $g_n$ are estimators of $\lambda$ and $g$. To simplify the notation, we define $\hat{\beta}_{m_{IPTW},\lambda}\equiv \hat{\beta}_n$. Under the positivity assumption, when $g_n$ and $\lambda_n$ are consistent, $\hat{\beta}_n$ is consistent and asymptotically linear (\citep[]{neugebauer2003locally}).  An important property of an asymptotically linear estimator is that it has a unique influence function (\citep[]{tsiatis2007semiparametric}). In other words, we identify $\hat{\beta}_n$ through its influence function. From the expression of the influence function we determine the asymptotic variance of $\hat{\beta}_n$ using Slutsky's theorem. Thus, we can construct the 95\% confidence interval for $\hat{\beta}_n$ given by:
\begin{equation}
\hat{\beta}_{jn} \pm 1.96 \times se(\hat{\beta}_{jn}),
\end{equation}
\noindent where $se(\hat{\beta}_{jn})=\hat{\sum}_{jn}^{p \times p}$ denotes the $j^{th}$ diagonal element of the sandwich variance estimator matrix of dimension $p\times p$ where $p$ number of parameters in $m(\bar{a},V|\beta)$. Figure (\ref{diagLCGMSM2}) provides a practical template of the estimation approach. In practice, when the distribution of $Y$ is part of the exponential family, the parameters of the working model can be directly estimated using the classical weighted \textit{generalized estimating equations} (GEE), with, for example, the function \textit{geeglm} from the \textbf{R} package \textit{geepack}. Indeed, under the causal assumptions described in Section \ref{sec2}, the estimating Equation (\ref{eq.est}) are solved by such a weighted GEE procedure.  When $Y$ is a time-to-event outcome, a weighted Cox model can be fitted using the \textbf{R} package \textbf{survival}. A simple practical solution for inference is to use the sandwich variance estimator of the GEE routine or of the Cox model, which treats the weights as known and results in conservative statistical inferences (\citep[]{hernan2010causal}). 

\newpage
\section{Simulation study} \label{sec6}
We aimed to evaluate the performance of our proposed approach for different types of outcomes, number of trajectory groups and follow-up times. We also wished to illustrate how alternative methods that make use of LCGM for causal inference can yield biased results in presence of time-dependent confounding. We have considered scenarios where: Y was a continuous, binary or time-to-event outcome; treatment and covariable were binary and time-varying. We chose $K=3,5$ and $10$ total number of follow-up times which yield respectively $8, 32$ and $1024$ possible treatment trajectories. These trajectories were summarized using LCGM in $J = $ 3, 4 and 5 trajectory groups. For all scenarios, we considered that the exposure and the covariate follow a binomial distribution (Equations are presented in the supplementary material, appendix B). 

To estimate the effect of the trajectory groups, we fitted a LCGM where the exposure $A$ is a linear function of the time on the logit scale. For each outcome, we estimated the working model $m(\bar{a},V|\beta)$ using either a weighted GEE regression or a weighted Cox regression with the inverse of probability of treatment as weights. In the case where $Y$ is continuous, we contrast our approach with a naive one where we adjusted for baseline and time-varying covariate when $K=3,5$.  This was not done for the case where $Y$ is binary or time-to-event outcome because of the problem of non-collapsibility. We also implemented an approach where after modeling the LCGM, probabilities of belonging in each trajectory groups are estimated by fitting a multinomial model with the trajectory group as dependent variable and the baseline characteristics of individuals as independents variables. In particular, \citep{lim2015all} estimated a MSM with the stabilized inverse of these probabilities as weights. They stabilized the weights with the marginal probability of the trajectory groups. We henceforth denote this approach as the inverse probability of trajectory groups weighting (IPTGW). We also applied this approach in the case where $Y$ was continuous. To measure the performance of the LCGM-MSM and alternatives approachs, we determined the true parameter using counterfactual data. Analytically determining the true value of the parameter of interest $\beta$ is challenging, particularly since this parameter is dependent on the chosen LCGM. The parameter of interest is thus susceptible to change from one simulation replicate to another. Instead, we determined the true value using a Monte Carlo simulation of counterfactual data (for more details see supplementary material, appendix B).


\section{Results}\label{sec7}

Table \ref{tab1} presents the LCGM-MSM and crude results for the scenarios with three trajectory groups. Similar results were obtained in the scenarios with four or five trajectory groups (see Web Table 1 and 2 in the online supplementary material, appendix C). The bias for LCGM-MSM when using either unstabilized or stabilized weights is small in all considered scenarios. However, stabilized weights yield estimators with a smaller standard deviation. The confidence intervals also tend to have better coverage when using stabilized weights in scenarios with multiple time points. Indeed, all confidence interval coverages were closed to 95\% (between 90\% and 98\%).  This is likely because stabilization tend to avoid extreme outlying weight values. On the other hand,  the crude model estimator has a large bias and low coverage probabilities of confidence intervals. Table \ref{tab2} presents the results for the baseline adjusted, time-varying adjusted and IPTGW adjusted LCGM. Recall that only the continuous outcome case, with either $K = 3$ or $K = 5$ time-points were considered for these alternatives methods. Moreover, regardless of the number of follow-up times and number of trajectory groups, alternative LCGM analyses were highly biased with low coverage of their confidence interval (between 2\% and 67\%).

\begin{table}[htbp]
  \centering
  \caption{LCGM-MSM with unstabilized and stabilized IPTW estimator when $Y$ is continuous, binary and time-to-eventt outcome for three trajectories.}
   \resizebox{0.9\textwidth}{!}{
    \begin{tabular}{lrrrrrrrrrrr}
    \toprule
          & \multicolumn{6}{c}{\textbf{Unstabilized IPTW}} & \multicolumn{5}{c}{\textbf{Stabilized IPTW}} \\
    \midrule
    \multicolumn{12}{c}{\textbf{$Y$ continous}} \\
    \midrule
    \midrule
    \multicolumn{12}{c}{\textbf{t=3}} \\
    \midrule
          & \multicolumn{1}{l}{Bias } & \multicolumn{1}{l}{SDE} & \multicolumn{1}{l}{CP } & \multicolumn{1}{l}{Crude bias} & \multicolumn{1}{l}{Crude CP } &       & \multicolumn{1}{l}{Bias } & \multicolumn{1}{l}{SDE} & \multicolumn{1}{l}{CP } & \multicolumn{1}{l}{Crude bias} & \multicolumn{1}{l}{Crude CP } \\
    Intercept & 0.010 & 0.200 & 95\%  & 0.040 & 6\%   &       & 0.000 & 0.100 & 95\%  & 0.150 & 11\% \\
    Group 2 & 0.000 & 0.480 & 95\%  & -0.200 & 16\%  &       & 0.000 & 0.370 & 95\%  & -0.250 & 31\% \\
    Group 3 & 0.000 & 0.200 & 96\%  & -0.340 & 8\%   &       & 0.000 & 0.160 & 97\%  & -0.370 & 0\% \\
    \midrule
    \multicolumn{12}{c}{\textbf{t=5}} \\
    \midrule
          & \multicolumn{1}{l}{Bias } & \multicolumn{1}{l}{SDE} & \multicolumn{1}{l}{CP } & \multicolumn{1}{l}{Crude bias} & \multicolumn{1}{l}{Crude CP } &       & \multicolumn{1}{l}{Bias } & \multicolumn{1}{l}{SDE} & \multicolumn{1}{l}{CP } & \multicolumn{1}{l}{Crude bias} & \multicolumn{1}{l}{Crude CP } \\
    Intercept & -0.010 & 0.370 & 92\%  & -0.090 & 36\%  &       & -0.010 & 0.210 & 94\%  & 0.170 & 17\% \\
    Group 2 & 0.000 & 0.770 & 93\%  & -0.050 & 19\%  &       & 0.000 & 0.520 & 95\%  & -0.120 & 50\% \\
    Group 3 & 0.010 & 0.360 & 94\%  & -0.420 & 12\%  &       & 0.030 & 0.230 & 95\%  & -0.350 & 1\% \\
    \midrule
    \multicolumn{12}{c}{\textbf{t=10}} \\
    \midrule
          & \multicolumn{1}{l}{Bias } & \multicolumn{1}{l}{SDE} & \multicolumn{1}{l}{CP } & \multicolumn{1}{l}{Crude bias} & \multicolumn{1}{l}{Crude CP } &       & \multicolumn{1}{l}{Bias } & \multicolumn{1}{l}{SDE} & \multicolumn{1}{l}{CP } & \multicolumn{1}{l}{Crude bias} & \multicolumn{1}{l}{Crude CP } \\
    Intercept & -0.024 & 0.308 & 87\%  & -0.132 & 39\%  &       & -0.036 & 0.200 & 86\%  & 0.130 & 28\% \\
    Group 2 & -0.008 & 0.638 & 90\%  & -0.287 & 25\%  &       & 0.027 & 0.431 & 92\%  & -0.137 & 50\% \\
    Group 3 & -0.037 & 0.375 & 88\%  & -0.497 & 5\%   &       & 0.026 & 0.268 & 91\%  & -0.352 & 1\% \\
    \midrule
    \multicolumn{12}{c}{\textbf{$Y$ binary}} \\
    \midrule
    \midrule
    \multicolumn{12}{c}{\textbf{t=3}} \\
    \midrule
          & \multicolumn{1}{l}{Bias } & \multicolumn{1}{l}{SDE} & \multicolumn{1}{l}{CP } & \multicolumn{1}{l}{Crude bias} & \multicolumn{1}{l}{Crude CP } &       & \multicolumn{1}{l}{Bias } & \multicolumn{1}{l}{SDE} & \multicolumn{1}{l}{CP } & \multicolumn{1}{l}{Crude bias} & \multicolumn{1}{l}{Crude CP } \\
    Intercept & 0.002 & 0.195 & 95\%  & 0.033 & 50\%  &       & -0.005 & 0.112 & 95\%  & 0.140 & 48\% \\
    Group 2 & 0.003 & 0.462 & 96\%  & -0.177 & 35\%  &       & 0.010 & 0.354 & 95\%  & -0.227 & 43\% \\
    Group 3 & 0.001 & 0.218 & 96\%  & -0.299 & 31\%  &       & 0.016 & 0.173 & 96\%  & -0.320 & 21\% \\
    \midrule
    \multicolumn{12}{c}{\textbf{t=5}} \\
    \midrule
          & \multicolumn{1}{l}{Bias } & \multicolumn{1}{l}{SDE} & \multicolumn{1}{l}{CP } & \multicolumn{1}{l}{Crude bias} & \multicolumn{1}{l}{Crude CP } &       & \multicolumn{1}{l}{Bias } & \multicolumn{1}{l}{SDE} & \multicolumn{1}{l}{CP } & \multicolumn{1}{l}{Crude bias} & \multicolumn{1}{l}{Crude CP } \\
    Intercept & -0.012 & 0.376 & 93\%  & -0.091 & 54\%  &       & -0.010 & 0.250 & 94\%  & 0.144 & 64\% \\
    Group 2 & 0.005 & 0.761 & 93\%  & -0.048 & 42\%  &       & 0.008 & 0.557 & 94\%  & -0.108 & 69\% \\
    Group 3 & 0.009 & 0.389 & 94\%  & -0.344 & 33\%  &       & 0.020 & 0.275 & 94\%  & -0.295 & 33\% \\
    \midrule
    \multicolumn{12}{c}{\textbf{t=10}} \\
    \midrule
          & \multicolumn{1}{l}{Bias } & \multicolumn{1}{l}{SDE} & \multicolumn{1}{l}{CP } & \multicolumn{1}{l}{Crude bias} & \multicolumn{1}{l}{Crude CP } &       & \multicolumn{1}{l}{Bias } & \multicolumn{1}{l}{SDE} & \multicolumn{1}{l}{CP } & \multicolumn{1}{l}{Crude bias} & \multicolumn{1}{l}{Crude CP } \\
    Intercept & -0.021 & 0.353 & 91\%  & -0.125 & 62\%  &       & -0.036 & 0.218 & 91\%  & 0.115 & 66\% \\
    Group 2 & 0.022 & 0.630 & 92\%  & -0.221 & 42\%  &       & 0.033 & 0.417 & 95\%  & -0.106 & 79\% \\
    Group 3 & -0.011 & 0.430 & 90\%  & -0.360 & 19\%  &       & 0.029 & 0.328 & 93\%  & -0.270 & 35\% \\
    \midrule
    \multicolumn{12}{c}{\textbf{$Y$ time-to-event }} \\
    \midrule
    \midrule
    \multicolumn{12}{c}{\textbf{t=3}} \\
    \midrule
          & \multicolumn{1}{l}{Bias } & \multicolumn{1}{l}{SDE} & \multicolumn{1}{l}{CP } & \multicolumn{1}{l}{Crude bias} & \multicolumn{1}{l}{Crude CP } &       & \multicolumn{1}{l}{Bias } & \multicolumn{1}{l}{SDE} & \multicolumn{1}{l}{CP } & \multicolumn{1}{l}{Crude bias} & \multicolumn{1}{l}{Crude CP } \\
    Group 2 & 0.000 & 0.33  & 96\%  & -0.111 & 25\%  &       & 0.003 & 0.274 & 95\%  & -0.155 & 37\% \\
    Group 3 & 0.003 & 0.14  & 96\%  & -0.186 & 25\%  &       & 0.002 & 0.140 & 95\%  & -0.240 & 4\% \\
    \midrule
    \multicolumn{12}{c}{\textbf{t=5}} \\
    \midrule
          & \multicolumn{1}{l}{Bias } & \multicolumn{1}{l}{SDE} & \multicolumn{1}{l}{CP } & \multicolumn{1}{l}{Crude bias} & \multicolumn{1}{l}{Crude CP } &       & \multicolumn{1}{l}{Bias } & \multicolumn{1}{l}{SDE} & \multicolumn{1}{l}{CP } & \multicolumn{1}{l}{Crude bias} & \multicolumn{1}{l}{Crude CP } \\
    Group 2 & 0.008 & 0.49  & 94\%  & -0.012 & 35\%  &       & 0.006 & 0.353 & 94\%  & -0.075 & 65\% \\
    Group 3 & 0.001 & 0.22  & 94\%  & -0.165 & 35\%  &       & 0.018 & 0.163 & 94\%  & -0.190 & 13\% \\
    \midrule
    \multicolumn{12}{c}{\textbf{t=10}} \\
    \midrule
          & \multicolumn{1}{l}{Bias } & \multicolumn{1}{l}{SDE} & \multicolumn{1}{l}{CP } & \multicolumn{1}{l}{Crude bias} & \multicolumn{1}{l}{Crude CP } &       & \multicolumn{1}{l}{Bias } & \multicolumn{1}{l}{SDE} & \multicolumn{1}{l}{CP } & \multicolumn{1}{l}{Crude bias} & \multicolumn{1}{l}{Crude CP } \\
    Group 2 & -0.004 & 0.34  & 91\%  & -0.100 & 40\%  &       & 0.020 & 0.284 & 96\%  & -0.069 & 85\% \\
    Group 3 & -0.016 & 0.18  & 91\%  & -0.158 & 35\%  &       & 0.029 & 0.181 & 91\%  & -0.166 & 52\% \\
    \bottomrule
    \end{tabular}}%
    
  Legend: bias, $CP$ and $SDE$ indicate respectively the bias, the coverage probability and the standard deviation of the estimate of $\beta_{IPTW}$ for the adjusted model with IPTW weight, crude \%bias and crude $CP$ indicate respectively the bias and the coverage probability of the estimate of $\beta_{crude}$ from the unadjusted model.
    \label{tab1}%
\end{table}%

\begin{table}[htbp]
  \centering
  \caption{LCGM-MSM with baseline adjustment, time varying covariates  adjustment and IPTGW as an estimation approach of a MSM when $Y$ is continuous.}
     \resizebox{0.85\textwidth}{!}{
    \begin{tabular}{lrrrrrrrrrrr}
    \toprule
          & \multicolumn{3}{c}{baseline adjustment} &       & \multicolumn{3}{c}{tvc adjustment} &       & \multicolumn{3}{c}{Stabilized IPTGW adjustment} \\
    \midrule
    \multicolumn{12}{c}{\textbf{3 trajectories}} \\
    \midrule
    \midrule
    \multicolumn{12}{c}{\textbf{K=3}} \\
    \midrule
          & \multicolumn{1}{l}{bias } & \multicolumn{1}{l}{SDE} & \multicolumn{1}{l}{CP} &       & \multicolumn{1}{l}{bias } & \multicolumn{1}{l}{SDE} & \multicolumn{1}{l}{CP} &       & \multicolumn{1}{l}{bias } & \multicolumn{1}{l}{SDE} & \multicolumn{1}{l}{CP} \\
    Intercept & -0.129 & 0.07  & 59\%  &       & -0.721 & 0.13  & 0\%   &       & 0.130 & 0.056 & 24\% \\
    Group 2 & -0.182 & 0.17  & 8\%   &       & -0.062 & 0.43  & 57\%  &       & -0.211 & 0.123 & 33\% \\
    Group 3 & -0.284 & 0.12  & 18\%  &       & -0.255 & 0.14  & 15\%  &       & -0.310 & 0.136 & 2\% \\
    \midrule
    \multicolumn{12}{c}{\textbf{K=5}} \\
    \midrule
          & \multicolumn{1}{l}{bias } & \multicolumn{1}{l}{SDE} & \multicolumn{1}{l}{CP} &       & \multicolumn{1}{l}{bias } & \multicolumn{1}{l}{SDE} & \multicolumn{1}{l}{CP} &       & \multicolumn{1}{l}{bias } & \multicolumn{1}{l}{SDE} & \multicolumn{1}{l}{CP} \\
    Intercept & -0.221 & 0.15  & 29\%  &       & -0.955 & 0.20  & 0\%   &       & 0.146 & 0.100 & 30\% \\
    Group 2 & -0.062 & 0.33  & 15\%  &       & -0.122 & 0.51  & 27\%  &       & -0.091 & 0.200 & 46\% \\
    Group 3 & -0.398 & 0.15  & 17\%  &       & -0.517 & 0.18  & 2\%   &       & -0.305 & 0.130 & 4\% \\
    \midrule
    \multicolumn{12}{c}{\textbf{4 trajectories}} \\
    \midrule
    \midrule
    \multicolumn{12}{c}{\textbf{K=3}} \\
    \midrule
          & \multicolumn{1}{l}{bias } & \multicolumn{1}{l}{SDE} & \multicolumn{1}{l}{CP} &       & \multicolumn{1}{l}{bias } & \multicolumn{1}{l}{SDE} & \multicolumn{1}{l}{CP} &       & \multicolumn{1}{l}{bias } & \multicolumn{1}{l}{SDE} & \multicolumn{1}{l}{CP} \\
    Intercept & -0.087 & 0.09  & 69\%  &       & -0.680 & 0.13  & 0\%   &       & 0.140 & 0.083 & 17\% \\
    Group 2 & -0.109 & 0.15  & 31\%  &       & -0.044 & 0.38  & 60\%  &       & -0.177 & 0.228 & 41\% \\
    Group 3 & -0.288 & 0.18  & 22\%  &       & -0.182 & 0.34  & 35\%  &       & -0.279 & 0.181 & 25\% \\
    Group 4 & -0.303 & 0.14  & 20\%  &       & -0.300 & 0.25  & 16\%  &       & -0.311 & 0.166 & 10\% \\
    \midrule
    \multicolumn{12}{c}{\textbf{K=5}} \\
    \midrule
          & \multicolumn{1}{l}{bias } & \multicolumn{1}{l}{SDE} & \multicolumn{1}{l}{CP} &       & \multicolumn{1}{l}{bias } & \multicolumn{1}{l}{SDE} & \multicolumn{1}{l}{CP} &       & \multicolumn{1}{l}{bias } & \multicolumn{1}{l}{SDE} & \multicolumn{1}{l}{CP} \\
    Intercept & -0.193 & 0.15  & 26\%  &       & -0.913 & 0.18  & 0\%   &       & 0.152 & 0.110 & 25\% \\
    Group 2 & 0.112 & 0.28  & 25\%  &       & -0.014 & 0.43  & 51\%  &       & -0.031 & 0.207 & 60\% \\
    Group 3 & -0.248 & 0.24  & 27\%  &       & -0.333 & 0.34  & 19\%  &       & -0.243 & 0.197 & 27\% \\
    Group 4 & -0.410 & 0.22  & 13\%  &       & -0.527 & 0.27  & 4\%   &       & -0.349 & 0.173 & 5\% \\
    \midrule
    \multicolumn{12}{c}{\textbf{5 trajectories}} \\
    \midrule
    \midrule
    \multicolumn{12}{c}{\textbf{K=3}} \\
    \midrule
          & \multicolumn{1}{l}{bias } & \multicolumn{1}{l}{SDE} & \multicolumn{1}{l}{CP} &       & \multicolumn{1}{l}{bias } & \multicolumn{1}{l}{SDE} & \multicolumn{1}{l}{CP} &       & \multicolumn{1}{l}{bias } & \multicolumn{1}{l}{SDE} & \multicolumn{1}{l}{CP} \\
    Intercept & -0.051 & 0.09  & 77\%  &       & -0.648 & 0.11  & 0\%   &       & 0.153 & 0.083 & 11\% \\
    Group 2 & -0.162 & 0.14  & 42\%  &       & -0.104 & 0.34  & 54\%  &       & -0.185 & 0.235 & 42\% \\
    Group 3 & -0.169 & 0.17  & 37\%  &       & -0.108 & 0.38  & 45\%  &       & -0.196 & 0.229 & 40\% \\
    Group 4 & -0.329 & 0.19  & 23\%  &       & -0.277 & 0.36  & 20\%  &       & -0.321 & 0.169 & 25\% \\
    Group 5 & -0.335 & 0.13  & 19\%  &       & -0.320 & 0.24  & 18\%  &       & -0.355 & 0.154 & 7\% \\
    \midrule
    \multicolumn{12}{c}{\textbf{K=5}} \\
    \midrule
          & \multicolumn{1}{l}{bias } & \multicolumn{1}{l}{SDE} & \multicolumn{1}{l}{CP} &       & \multicolumn{1}{l}{bias } & \multicolumn{1}{l}{SDE} & \multicolumn{1}{l}{CP} &       &       &       &  \\
    Intercept & -0.178 & 0.15  & 28\%  &       & -0.892 & 0.18  & 0\%   &       & 0.159 & 0.110 & 21\% \\
    Group 2 & 0.127 & 0.25  & 39\%  &       & -0.002 & 0.36  & 57\%  &       & -0.031 & 0.223 & 67\% \\
    Group 3 & -0.104 & 0.28  & 41\%  &       & -0.154 & 0.39  & 42\%  &       & -0.157 & 0.243 & 50\% \\
    Group 4 & -0.264 & 0.23  & 31\%  &       & -0.368 & 0.34  & 19\%  &       & -0.280 & 0.211 & 27\% \\
    Group 5 & -0.424 & 0.29  & 14\%  &       & -0.555 & 0.31  & 5\%   &       & -0.390 & 0.185 & 4\% \\
    \bottomrule
    \end{tabular}}
    
 Legend: bias, $CP$ and $SDE$ indicate respectively the bias, the coverage probability and the standard deviation of the estimate of $\beta$ for each adjustment approach. Abbreviations : tvc=time-varying covariates, IPTGW= inverse of probability of trajectory groups weights.
  \label{tab2}%
\end{table}%

\clearpage
\section{Discussion}\label{sec9}
We proposed a new approach combining LCGM and MSMs. 
In a first step, we propose to use LCGM to classify individuals into a few latent classes based on their treatment adherence pattern, then in a second step, choose a working MSM that relates the outcome to these latent classes.   The parameter of interest is nonparametrically defined as the projection of the true MSM onto the chosen working causal model. This approach has a double benefits. On one hand, we avoid assuming that we know the functional form of the true model as in a parametric model. On another hand, we also avoid the curse of dimensionality that is often encountered when estimating a nonparametric model with finite sample data.  We showed that, the data-driven estimation of the trajectory groups can be ignored. As such, parameters can be estimated using the IPTW estimator and conservative inferences can be obtained using a standard robust variance estimator. Simulation studies were used to illustrate our approach and compare it with unadjusted, baseline covariates-adjusted, time-varying covariates adjusted and inverse probability of trajectory groups weighting adjusted alternatives. Scenarios with varying types of outcomes (continuous, binary or time-to-event) and number of follow-up times (3, 5, 10) and number of trajectory classes (3, 4, 5) were considered. In the simulation study, we found that our proposed approach yielded estimators with little or no bias. Alternative analyses were biased and had low coverage of their confidence intervals.  

Despite the strengths of our proposal, some limitations need to considered. For example, in presence of time-varying confounding, the local independence assumption of the LCGM is unlikely to hold. Indeed, when treatment uses at different time-points are locally dependent, overlapping or redundant information is accounted for when it should not be in the estimation of the LCGM (\citep[]{vermunt2002latent}). Hence, omitting time-varying covariates when modeling trajectory groups might lead to an overestimation of the ``posterior'' probability.   Also, we chose arbitrarily the number of trajectory groups in the simulation study and in the application. As \citep{nagin2005group} mentioned the groups can be seen as points of support on the unknown distribution of trajectories. In that sense, the groups are not meant to represent the true data-generating process.  In the case when $Y$ is binary, one might be cautious regarding the number of available events per trajectory group to avoid estimation problems.  

Our proposed approach can be improved by using a doubly robust estimator instead of the IPTW estimator. It is clear in our simulation results that as the number of follow-up time increased, the IPTW estimator failed to completely eliminate the bias. Indeed, the IPTW estimator heavily relies on a correct specification of the probability of treatment assignment and is sensitive to near violations of the positivity assumptions. Furthermore, relaxing the assumption of local independence might also improve the LCGM-MSM through a better classification of individuals' treatment patterns. The number of trajectory groups can be chose using the usual recommendations in the literature such as the BIC criterion or by cross-validation (\citep[]{nagin2005group, nielsen2014group}). 
In conclusion, we believe that our LCGM-MSM can be useful to correctly estimate the effect of trajectory groups on an outcome when conducting longitudinal studies based on administrative data.  Our proposed approach might help practitioners to have a better idea of what type of adherence profile is worth targeting as LCGM is better to measure treatment adherence pattern than most other commonly used methods.

\section*{Acknowledgments}
This project was funded by a CIHR grant [$\#$ 165942] and a start-up award from  Facult\'e de M\'edecine de l'Universit\'e Laval to DT. CS, JRG and DT are supported by career awards from the Fonds de recherche du Qu\'ebec -- Sant\'e. AD received a scholarship from VITAM." 

\bibliographystyle{biorefs}
\bibliography{refs_fp}

\end{document}